\documentstyle[12pt,twoside,fleqn,espcrc1]{article}
\newcommand{\AmS}{{\protect\the\textfont2
  A\kern-.1667em\lower.5ex\hbox{M}\kern-.125emS}}

\title {\null\vspace*{-3.0cm}\hfill {\small ORNL-CTP-9803 and
hep-ph/9808402} \\ \vskip 0.8cm Unresolved Questions in $J/\psi$
Production and Propagation in Nuclei}

\author{Cheuk-Yin Wong \address{Physics Division, Oak Ridge National
Laboratory, Oak Ridge, TN 37831 }
}
\begin{document}
\def\bbox{ \vec}
\maketitle

\begin{abstract}
{In order to understand the $J/\psi$ suppression arising from the
possible occurrence of the quark-gluon plasma in high-energy heavy-ion
collisions, it is necessary to have a comprehensive picture how the
$J/\psi$ and its precursors are produced, what their properties after
production are, and how the $J/\psi$ and its precursors propagate
inside nuclear matter.  There are unresolved questions in the
descriptions of $J/\psi$ production and propagation.  We outline some
of these questions and discuss the approaches for their resolution.  }

\end{abstract}

\vspace*{-0.3cm}
\section{Introduction}
\vspace*{-0.3cm}

The occurrence of $J/\psi$ suppression has been suggested as a way to
probe the screening between a charm quark-antiquark pair in the
quark-gluon plasma \cite{Mat86}.  With the observation of deviation
from systematics in the absorption of $J/\psi$ in high-energy
nucleus-nucleus Pb-Pb collisions \cite{Gon96,Lou96}, it is necessary
to understand the complete process of $J/\psi$ (and $\psi'$)
production and propagation in hadron matter in order to infer the
interesting physics of their propagation in the deconfined quark-gluon
plasma.

While much progress has been made in many aspects of this topic
\cite{Ein75,Fri77,Cha78,Bod95,Bra96,Ben96,Cac97b,Pet98,Kha96,Won96,Won97b,Won98a},
unresolved problems remain.  These problems do not exist in isolation,
as the suppression of the $J/\psi$ depends on how the produced $c\bar
c$ object interacts with hadrons and deconfined matter. This
interaction depends on the properties of the produced $c\bar c$, which
in turn depend on the production mechanism.

Because of the limited space in this survey, we shall focus on the
unresolved problems in the production and propagation of the $J/\psi$
and its precursors in nucleon-nucleon and nucleon-nucleus collisions.
$J/\psi$ suppression in nucleus-nucleus collisions has been discussed
earlier in \cite{Kha96,Won96,Won97b,Won98a} and will be the subject of
another report.

\vspace*{-0.4cm}
\section{  The Production of $c\bar c$ Bound States in a
Nucleon-Nucleon Collision}
\vspace*{-0.3cm}

In a nucleon-nucleon collision, the parton of one nucleon collides with
the parton of another nucleon.  There is a finite probability for the
collision of two gluons or a $q$ and a $\bar q$ to produce a $c\bar c$
pair.  We focus our attention on those $c\bar c$ pairs which have
center-of-mass energies close to the bound states energies and can
form various $c\bar c$ bound states.

The simplest theory for heavy quarkonium production is the
Color-Evaporation Model \cite{Ein75,Fri77}.  The model is a
prescription which calculates first the total cross section for $c\bar
c$ pair production up to the $D\bar D$ threshold.  The cross section
for $J/\psi$ production is then obtained empirically as a fraction of
this total cross section.  Although the Color-Evaporation Model is
useful in providing a rough estimate of the quarkonium production
cross section, how the colored $c\bar c$ system evolves into a
color-singlet $J/\psi$ state is not specified.

The Color-Singlet Model (CSM) \cite{Cha78} was developed to examine the
production mechanism in more detail.  It is assumed that the
production probability amplitude is factorizable and is the product of
a short-distance perturbative QCD part and a long-distance
nonperturbative part.  The PQCD part is given in terms of the Feynman
production amplitude.  The long-distance nonperturbative part is given
in terms of the wave function (for $J/\psi$ production) or its
derivative (for $\chi$ production) at the origin, to take into account
the interaction of the final $c$ and $\bar c$ quarks. The
color-singlet component of the production amplitude is projected out,
to be identified as the probability amplitude for the production of
the observed bound states.

As the experimental data of heavy quarkonium production accumulates,
the CSM has been found to be inadequate in describing the production
process (see \cite{Bod95,Bra96,Ben96,Cac97b} for reviews).  First, the
model predicts cross sections for high-$p_T$ $J/\psi$, $\psi'$ and
$\chi$ production for $p$+$\bar p$ collisions at the Tevatron at
$\sqrt{s}=1.8$ TeV which are substantially lower than the cross
sections measured by the CDF Collaboration \cite{Abe97}.  Second, the
CSM leads to a cross section ratio $\chi_1/\chi_2$ equal to 0.067
\cite{Van95} for $\pi$-$N$ collision at 300 GeV, which is too small
compared to the experimental data of $\sim$0.70 \cite{Ant93}.  Third,
CSM predicts $J/\psi$ production with a population of $J_z=\pm 1$
states substantially greater than that of the $J_z=0$ state, leading
to an anisotropic angular distribution of muons from $J/\psi$ decay.
The angular distribution is often parameterized in the form of
$1+\lambda \cos^2 \theta$ in the Gottfried-Jackson frame, which is the
$J/\psi$ rest frame in which the beam goes in the $z$ directions,
while the target momentum lies in the $x$-$z$ plane with momentum
pointing downward ($p_x \leq 0)$. The CSM gives $\lambda \sim 0.5$,
but the experimental data give an isotropic distribution with
$\lambda\sim 0$ for $J/\psi$ and $\psi'$ production in $\pi$-W
collisions at 252 GeV \cite{Bin87} and 125 GeV \cite{Ake93}.

The Color-Octet Model (COM) was developed to include processes which
are additional to those in the color-singlet model \cite{Bod95}.  It
is assumed that besides the production of the bound state by the
color-single mechanism, bound states are also produced by the
color-octet formalism whereby a $c\bar c$ pair in a color-octet state
is first formed either by gluon fragmentation or by direct parton
reactions, and the octet color of the $(c\bar c)_8$ pair is
neutralized by emitting a soft gluon of low energy and momentum.  The
emission of the soft gluon takes place at a nonperturbative QCD time
scale and is assumed to occur with a large probability.  The cross
sections for direct $J/\psi$ and $\psi'$ production due to the
color-octet mechanism are then proportional to a lower power of
$\alpha_s$, leading to large color-octet contributions, in comparison
with the higher-order process of color-singlet $J/\psi$ and $\psi'$
production by hard gluon emission.  The matrix elements for the
emission of the soft gluon from the color-octet $(c\bar c)_8$ states
involve nonperturbative QCD.  They are left as phenomenological
parameters obtained by fitting experimental data.  Matrix elements
have been extracted to yield good agreement with the CDF data
\cite{Cho96}.

\vspace*{-0.6cm}
\section{  Unresolved Questions in the Color-Octet Model}
\vspace*{-0.3cm}

Although the CDF data for high $p_T$ quarkonium production can be
understood in terms of an additional color-octet mechanism, there are
many unresolved questions in the Color-Octet Model.  The first
question is connected with $J/\psi$ production in $\gamma p$ reactions
at energies of 40 GeV $<\sqrt{s_{\gamma p}}<$ 140 GeV at HERA
\cite{Cac97a}. One can examine the transverse momentum $p_T$ and the
momentum fraction $z$ of the produced $J/\psi$, where $z=(p\cdot
p_{J/\psi})/(p\cdot p_\gamma)$ and $p$ is the momentum of the proton.
Because the photon is in a color-singlet state, the fusion of the
photon with a gluon leads to a color-octet state, which according to
the COM can evolve into a color-singlet state by the emission of a
soft gluon of little energy and momentum.  On the other hand, direct
color-singlet production will involve the emission of a hard gluon
through the reaction $\gamma+g \rightarrow J/\psi+ g$.  Because a
produced hard gluon carries transverse momentum and energy while the
initial colliding $\gamma$ and gluon carry little transverse momentum,
the kinematic region of $p_{T}<$1 GeV and another region with $z\sim$1
are regions where the color-octet production process dominates.  The
color-singlet production process dominates in the other regions with
$p_{T} >$ 1 GeV and $z< 1$.  However, using the color-octet matrix
element as determined from the CDF measurements and assuming $\langle
{\cal O} _8 ^{J/\psi} ({}^1 S_0) \rangle$=$\langle {\cal O} _8
^{J/\psi} ({}^3 P_0) \rangle/m_c^2$, one obtains theoretical cross
sections which are much larger than the experimental data for $p_T <
1$ GeV and for $z\sim 1$ \cite{Cac97a}.

The second question is connected with the color-octet matrix elements
extracted from $J/\psi$, $\psi'$ and $\chi$ production cross sections
at fixed-target energies which give \cite{Ben96}
\begin{eqnarray}
\label{eq:1}
\langle {\cal O} _8 ^{J/\psi} ({}^1 S_0) \rangle + {7 \over m_c^2} 
\langle {\cal O} _8 ^{J/\psi} ({}^3 P_0) \rangle = 3.0 \times 10^{-2}
{\rm~ GeV}^3
\end{eqnarray}
for $J/\psi$ production, and for $\psi'$ production
\begin{eqnarray}
\label{eq:2}
\langle {\cal O} _8 ^{\psi'} ({}^1 S_0) \rangle + {7 \over m_c^2} 
\langle {\cal O} _8 ^{\psi'} ({}^3 P_0) \rangle = 0.5 \times 10^{-2}
{\rm~ GeV}^3.
\end{eqnarray}
These need to be compared with those from high $p_T$ CDF measurements
at 1.8 TeV \cite{Cho96}
\begin{eqnarray}
\langle {\cal O} _8 ^{J/\psi} ({}^1 S_0) \rangle + {3 \over m_c^2} 
\langle {\cal O} _8 ^{J/\psi} ({}^3 P_0) \rangle = 6.6 \times 10^{-2}
{\rm~ GeV}^3
\end{eqnarray}
for $J/\psi$ production, and for $\psi'$ production
\begin{eqnarray}
\label{eq:del8}
\langle {\cal O} _8 ^{\psi'} ({}^1 S_0) \rangle + {3 \over m_c^2}
\langle {\cal O} _8 ^{\psi'} ({}^3 P_0) \rangle = 1.8 \times 10^{-2}
{\rm~ GeV}^3.
\end{eqnarray}
As mentioned by Beneke $et~al.$ \cite{Ben96}, the fixed target energy
values are a factor of 4 (7) smaller than the Tevatron values for
$J/\psi$($\psi'$) if $\langle {\cal O} _8 ^{ \{\psi(\psi') \}} ({}^1
S_0) \rangle = \langle {\cal O} _8 ^{ \{\psi(\psi') \}} ({}^3 P_0)
\rangle/m_c^2$.  The discrepancy would be lower to a factor of 2 if
$\langle {\cal O} _8 ^{\{\psi(\psi')\}} ({}^3 P_0) \rangle=0$.

The third question which is left from the Color-Singlet Model concerns
$\chi_1$ production relative to $\chi_2$ production.  As we mentioned
earlier, the color-singlet model prediction of the $\chi_1$ cross
section is much smaller than the experimental data \cite{Van95} when
we normalize the theoretical yield to $\chi_2$ states.  The
Color-Octet Model does not improve this discrepancy on the $\chi_1$
yield.  This arises because of Yang's theorem that two on-shell gluons
cannot fuse to form a spin-one state.  Hence, $\chi_1$ can be formed
only by the $gg \rightarrow \chi_1 g$ process which is $\alpha_s^3$
order in cross section.  On the other hand, $\chi_2$ in the
color-singlet state can be formed by the fusion of two gluons with a
cross section that is of order $\alpha_s^2$.  Thus, the predicted
${\cal B}_1\chi_1/{\cal B}_2\chi_2 = 0.13 $ \cite{Ben96} is much
smaller than the observed ${\cal B}_1\chi_1/{\cal B}_2\chi_2 \sim 1.4$
for $\pi$-$N$ collision at 185 GeV and 300 GeV \cite{Ant93}.

Finally, previous questions on the polarization of the produced
$J/\psi$ and $\psi'$ for the Color-Singlet Model remains a puzzle for
the Color-Octet Model.  In the Color-Octet Model, the dominant
production arises from the $c\bar c$ color-octet pair in the ${}^1S_0$
and the ${}^3 P_J$ states whose matrix elements are in the combination
of Eqs.\ ({\ref{eq:1}) and ({\ref{eq:2}).  While the ${}^1S_0
\rightarrow {}^3S_1$ transition leads to an isotropic muon
distribution, the ${}^3 P \rightarrow {}^3S_1$ preferentially
populates $J_z=\pm 1$ substates with large transverse polarization.
The velocity counting rule gives $\langle {\cal O} _8 ^{\psi'} ({}^1
S_0) \rangle$ to be of the same order as $\langle {\cal O} _8
^{J/\psi} ({}^3 P_0) \rangle/m_c^2$.  If so, the expected angular
distribution is not isotropic \cite{Ben96}.  The experimental
polarization suggests a nearly isotropical distribution in its
center-of-mass frame as would be the case for equal population of the
three $J_z$ substates.

\vspace*{-0.4cm}
\section{  Resolution of Some of the Questions in $J/\psi$ Production}
\vspace*{-0.3cm}

We can discuss possible approaches which may resolve some of these
questions.  First, we examine the question of $J/\psi$ polarization.
The polarization data can be explained as arising from the fact that
in the color-octet mechanism, the soft gluon is emitted predominantly
from a ${}^1S_0$ state to reach the final state $J/\psi$ state of
${}^3S_1$ -- much more prominent than from the emission from the
${}^3P_J$ state. This may appear surprising as one expects the
opposite relation from the velocity counting rule, where from the
comparison of the spatial transition currents, the absolute value of
the ratio of the M1 amplitude to that of the E1 amplitude is of the
order of $v$ \cite{Bla52}.  However, the transition matrix elements
consist of a part due to the spatial current and a part due to the
spin current. They have different dependences on the energy of the
radiation.  The transition from ${}^3P_0$ indeed dominates over the
${}^3S_1$ transition for high energy radiative transitions, but the
situation is just reversed for soft radiative transitions when the
transition due to the spin current is allowed, as was clearly
demonstrated by the analogous situation in the radiative production of
deuterons by the interaction of a neutron with a proton \cite{Bla52}.

The deuteron bound state is a ${}^3S_1$ state, with the same quantum
number as $J/\psi$ and $\psi'$.  A low energy neutron and a proton can
form ${}^1S_0$ and ${}^3P_J$ states in the continuum.  The cross
section for a radiative transition from the ${}^1S_0$ continuum state
to the bound ${}^3 S_1$ deuteron state with the emission of a photon is
[Eq. (XII.4.27) of \cite{Bla52}]
\begin{eqnarray}
\label{eq:M1}
\sigma^{M1}({}^1S_0 \rightarrow {}^3S_1) = \pi {e^2 \over \hbar c}
\biggl ( {\hbar \over M_N c}\biggr)^2 { B \over M c^2} (\mu_n - \mu_p)^2
(1-\gamma a_s) \biggl ( {2B \over E_N } \biggr )^{1/2},
\end{eqnarray}
where $B \equiv \hbar^2 \gamma^2/M_N$ is the binding energy of the
deuteron, $E_N \equiv 2\hbar^2 k^2/M_N=2(E_{\rm photon} - B)$ is the
asymptotic neutron kinetic energy relative to the proton, $\mu_n$ and
$\mu_p$ are the magnetic moments of the neutron and proton
respectively, and $a_s$ is the scattering length between the neutron
and proton.  The M1 cross section has the typical
$1/v=1/\sqrt{2E_N/M_N}$ behavior, and is large for very soft radiative
transitions.  The magnetic dipole transition arises not from the
spatial current but from the spin current which allows a spin-flip in
the ${}^1S_0 \rightarrow {}^3 S_1$ transition.

On the other hand, the E1 radiative transition from a ${}^3 P_J$
continuum state to the bound ${}^3 S_1$ deuteron state with the
emission of a photon comes from the spatial transition current.  The
radiative transition cross section is [Eq. (XII.4.38) and (XII.4.14) of
\cite{Bla52}]
\begin{eqnarray}
\label{eq:E1}
\sigma^{E1}({}^3 P_J \rightarrow {}^3S_1) =
{8 \pi \over 3} {e^2 \over \hbar c}
{k \gamma \over (k^2+\gamma^2)^2 }
(1-\gamma r_{ot})^{-1} ,
\end{eqnarray}
where $r_{ot}$ is the effective range of the interaction.  From these
results, we note that for the softest radiation near the bound state
formation threshold, the magnetic dipole M1 cross section
$\sigma^{M1}$ varies as $1/v$ and is large, while the electric dipole
$E1$ cross section $\sigma^{E1}$ vanishes as $M_N v/2$=$\hbar
k\rightarrow 0$.  Thus, radiative formation of ${}^3S_1$ by the
emission of very soft radiation is dominated by the magnetic dipole
transition ${}^1 S_0 \rightarrow {}^3S_1$ over the electric dipole
transition ${}^3 P_J \rightarrow {}^3S_1$, and the velocity counting
rule breaks down \cite{Bla52}.  Experimentally, the dominance of the
${}^1S_0 \rightarrow {}^3 S_1$ transition over the ${}^1P_J
\rightarrow {}^3 S_1$ transition in the soft photon region has been
demonstrated in nuclear reactions \cite{Bla52}.

We can examine the color-octet matrix element by making the analogy of
photon radiation with gluon radiation.  For $c$ and $\bar c$ in
color-octet states, the interaction is repulsive, and $a_s$ is
negative.  It is clear from the deuteron analysis that for $J/\psi
({}^3 S_1)$ production by very soft gluon radiation from a color-octet
state, the ${}^1S_0 \rightarrow {}^3S_1$ transition dominates over
the ${}^1P_J \rightarrow {}^3 S_1$ transition.  With the dominance of
the ${}^1S_0\rightarrow {}^3S_1$, the resultant population of the
${}^3S_1$ states is unpolarized, with isotropic distribution of muons
from its decay.  This explains the observed isotropic muon angular
distribution.

With the dominance of $\langle {\cal O}_8^{J/\psi}({}^1S_0)\rangle$
over $\langle {\cal O}_8^{J/\psi}({}^3 P_J)\rangle/m_c^2$, the
discrepancy of the matrix elements in the Tevatron measurements and the
fixed target measurements is reduced.  The discrepancy can be further
reduced when one takes into account the physical masses and allowing
for the finite energy carried by the soft gluon \cite{Won98b}.  If
$\langle {\cal O}_8^{J/\psi}({}^3 P_J)\rangle$ is set to zero for soft
gluon radiation, then the CDF data imply $\langle {\cal
O}_8^{J/\psi}({}^1S_0)\rangle = 6.6 \times 10^{-2} {\rm ~~GeV}^2$, and
the HERA data of $\gamma p \rightarrow J/\psi X$ with $z\le 0.8$ and
$p_T \ge 1$ GeV remain approximately consistent with theoretical
predictions.  Howvere, the discrepancy at $z\sim 1$ remains and may
suggest additional effects at $z\sim 1$ which may be beyond the scope
of the COM.

Finally, the question of the ratio $\chi_1/\chi_2$ is not yet
resolved.  It has been suggested that such a discrepancy may be due to
higher twist terms \cite{Ben96,Tan96}.  One can think of another
intriguing possibility which may lead to an enhanced production of
$\chi_1$.  Experimentally, one knows that feeding from higher states
can be the source of observed bound state populations.  For example,
about 30.5\% of the observed $J/\psi$ comes from the feeding from the
electromagnetic decay of the $\chi$ states, most notably the $\chi_1$
and the $\chi_2$ states, and about 7.5\% from the decay of the $\psi'$
state \cite{Ant93}.  It is therefore of interest to examine whether
the $\chi_1$ states come from the feeding from higher $D$ states.  $D$
states form the multiplet $^1D_2$, and $^3D_{\{1,2,3\}}$.
Experimentally, the $J$=1 ${}^3D_1(1^{--})$ state has been observed
and lies at 3.77 GeV, just above the $D\bar D$ threshold of 3.74 GeV.
It decays predominantly by $D\bar D$ breakup.  The ${}^{\{ 1,3 \}}D_2$
and $^3D_3$ states have not been observed, but are expected to lie
close to the ${}^3D_1(1^{--})$ state.  They are likely to lie below
the $\pi D\bar D$ threshold.  While the $J$=3 $^3D_3(3^{--})$ state
can decay by $D \bar D$ breakup, the $J$=2 $^1D_2(2^{-+})$ and
$^3D_2(2^{--})$ states cannot decay by $D\bar D$ breakup because they
are unnatural parity states with $J^P$=$2^-$.  The $^3D_2(2^{--})$
state can predominantly decay into the $\chi_1$ and $\chi_2$ states by
E1 electromagnetic transitions \cite{Bar93}.  Such a state can feed
into the population of $\chi_1$ and $\chi_2$.  On the other hand,
${}^3D_2(2^{--})$ can be formed by the fusion of two gluons into a
color-singlet state with a cross section of order $\alpha_s^2$.  The
production of ${}^3D_2(2^{--})$ may result in an enhanced production
of $\chi_1$ and $\chi_2$, and may explain the discrepancy with regard
to the $\chi_1$ yield in the COM.

One concludes from the above discussions that while there are
questions concerning the Color-Octet Model, there may be resolution of
these questions in terms of a careful refinement on the details of the
model.

\vspace*{-0.4cm}
\section{  Propagation of $J/\psi$ Precursors in Nuclear Matter}
\vspace*{-0.3cm}

The COM gives the probability for the formation of various bound
states, but gives no specific information on the type of admixture of
the precursor state.  However, the nature of the state vector of the
precursor has great influence in its subsequent interaction with
target nucleons when the precursor propagates through nuclear matter
in nucleon-nucleus collisions.  Therefore, nucleon-nucleus collisions
provide an arena to study the nature of the state vector of the
precursor. The precursor state has many degrees of freedom: color $C$,
angular momentum and spin $JLS$.  The precursor state can, in general,
be incoherent in one degree of freedom but a coherent admixture in
another degree of freedom.  Theoretical and experimental
investigations of the nature of the admixture in the precursor state
are interesting and unresolved problems in $J/\psi$ production.

Dynamical processes in $J/\psi$ production are controlled by the
$c\bar c$ pair production time $\tau_{\rm pair}$ and the evolution
time $\tau_{\rm evol}$ for the produced $c\bar c$ pair to evolve into
a bound state.  The pair production is a short-distance perturbative
QCD process; its time $\tau_{\rm pair}$ is about $1/2m_c$=0.07 fm/c.
The evolution time is a nonperturbative QCD process, and its time
$\tau_{\rm evol}$ is about $1/\Lambda_{QCD}$=0.5 fm/c measured in the
$J/\psi$ rest frame.  In $J/\psi$ production in $pA$ collisions, the
dynamics is further controlled by the next-nucleon meeting time
$\tau_{\rm nn}$, the time it takes for the $J/\psi$ precursor to meet
the next target nucleon after its production.  In the $J/\psi$ rest
frame, the next-nucleon meeting time $\tau_{\rm nn}$ is $d/(\gamma^2
-1)^{1/2}$, where $d=2$ fm is the internucleon spacing in a nucleus at
rest and $\gamma (x_{{}_F})$ is the relativistic energy/mass ratio of
the moving target nucleons \cite{Won98a}.  For $x_{{}_F}>0$ at
fixed-target energies of several hundred GeV, we have $\tau_{\rm nn}
<< \tau_{\rm evol} $.  Therefore, for $p$-$A$ collisions in
fixed-target experiments, many of the collisions between the produced
precursor and target nucleons take place before the $c\bar c$ pair has
completed its evolution to bound color-singlet states.  One can use
nuclear matter in a $pA$ collision as an arena to probe the nature of
the $CJLS$ admixture of the produced precursor.

\vspace*{-0.4cm}
\section{ Color-Dependence of the Absorption Cross Section}
\vspace*{-0.3cm}

To study the absorption of $J/\psi$ in $pA$ collisions, we can rely
conceptually on the fact that the collision between the precursor and
the target nucleon occurs at high energies and high-energy
hadron-hadron cross sections are dominated by Pomeron exchange.  In
the Two-Gluon Model of the Pomeron (TGMP) \cite{Low75,Dol92,Woncw96b},
the color-singlet (C1) total $(c\bar c)_1$-nucleon cross $\sigma_1$
can be expressed as $T_1 - T_2$, where $T_n$ is the contribution in
which the two exchanged gluons interact with $n$ quarks in the
projectile. The color-singlet total $(c\bar c)_1$-$N$ cross sections
are size-dependent.  The cross section $\sigma_1$ vanishes if one of
the colliding hadrons shrinks to a point, because in this limit $T_2 =
T_1$.  A produced coherent color-singlet wave packet with a small
separation between $c$ and $\bar c$ will lead to small $J/\psi$
absorption, while a large separation in the wave packet will result in
a large absorption.

For $(c\bar c)_8$-$N$ scattering, the total color-octet (C8) cross
section $\sigma_8$ is very different, as pointed out by Dolej\v s\' i
and H\"ufner \cite{Dol92}. This is because the one- and two-quark
contributions now add together in the form of $T_1 + T_2/8$.  The
result is then insensitive to the $c$-$\bar c$ separation of the
color-octet precursors. It is also very large, typically of the order
of 30-60 mb when a perturbative propagator is used for gluons with a
nonzero effective mass. Because of the insensitivity of the cross
section on the color separation, the absorption cross section will be
insensitive to the spatial admixture of the color-octet precursor.

For a precursor with a coherent admixture of color-singlet and
color-octet states in the form ($a_1~{\rm C1} + a_2~ {\rm C8}$) the
total cross section between a precursor with nucleons in the TGMP is
approximately $|a_1|^2 \sigma_1 + |a_2|^2 \sigma_8$ where
$\sigma_{\{1,8\}}$ are the total cross sections evaluated with a pure
color-singlet or octet state \cite{Won97b}.  Thus, the cross section
for an incoherent admixture lies in between the two limits.

\vspace*{-0.4cm}
\section{  Incoherent Admixture } 
\vspace*{-0.3cm}

The simplest description of the precursor is the model of incoherent
admixture of color and spatial quantum numbers.  We must now
generalize this standard absorption model for precursors of one type
\cite{And77} to several types of precursors \cite{Won98a}.  The
survival probability is then the weighted sum of survival
probabilities of different absorption components characterized by
different absorption cross sections $\sigma_i$
\begin{eqnarray}
\label{eq:ic}
R (BA/NN, x_{{}_F}) && \equiv 
{d\sigma_{J/\psi}^{AB}/dx_F \over A d\sigma_{J/\psi}^{NN}/dx_F}
= \sum_{i} f_i (x_{{}_F}) 
R_{i}(AB)\,,  
\end{eqnarray}
where the production fractions are normalized to $\sum_i f_i = 1$,
$i=\{CJLS\}$, and
\begin{eqnarray}
\label{eq:xsec}
R_{ i}(BA) = \int {d{\bf b}_A \over A\sigma_{{\rm abs} \, i}} {d{\bf
b}_B \over B\sigma_{{\rm abs} \, i} } \biggl \{ 1 - \biggl ( 1- T_B
({\bf b}_B) \sigma_{{\rm abs} \,i} \biggr )^B \biggr \} \biggl
\{ 1 - \biggl ( 1- T_A ({\bf b}_A) \sigma_{{\rm abs} \, i} \biggr )^A
\biggr \} \, .
\end{eqnarray}
Each survival probability is approximately an exponential function of
the average path length $L$.  In the semilog plot of the logarithm of the
total survival probability as a function of the average path length,
the slope of the curve will change as a function of the average path
length.  The slope will be proportional to the largest absorption
cross section for small $L$, and to the smallest absorption cross
section for large $L$.

Ref. \cite{Won98a} considers the incoherent model in which the C1 and
C8 cross sections can be considered to be constants, independent of
the angular momentum $JLS$ and the number of nodes $n$. Then there are
only two components in Eq.\ (\ref{eq:ic}).  An absorption model can be
constructed that respects the popular theoretical prejudices that C1
precursors are produced in point-like states and tend to be transparent
with $\sigma_1$=0 in the colliding nuclear complex, where C8
precursors are strongly absorbed, with $\sigma_8\sim$15 mb. The C8
fractions that come out of this model are quite substantial, in
agreement with independent analyses of hadron production rates in free
space \cite{Ben96,Tan96}.  These results are supported by later
investigations \cite{Qia97}.  However, the phenomenological analyses
seem to show that the available data alone are not sufficiently
discriminating to tell us if the C1 precursors are transparent because
they are produced point-like, or if they are also absorbed because they
are produced at almost full size.  Better fits to these data are
obtained by using an older picture in which color-singlet precursors are
also absorbed by nuclei \cite{Won98a}.

\vspace*{-0.4cm}
\section{  Coherent Admixture } 
\vspace*{-0.3cm}

Experimental quarkonium production data have been obtained only for a
very limited number of nuclear mass numbers, which makes it difficult
to determine with certainty whether the survival probability as a
function of average path lengths has one or many absorption
components.  Alternatively fits by a single absorption component can
be made when one makes allowance for the energy dependence of the
production cross section \cite{Gon96,Lou96}.  Such a procedure has a
high degree of uncertainty because of the uncertainty in matching data
from different experimental conditions and energies.  It is an
unresolved problem as to whether the survival probability can be described
by a single exponential component or the sum of many components.

A description of the precursor as a coherent admixture has been given
previously \cite{Won97b}.  In this description, the interaction of
partons $a$ and $b$ in a nucleon-nucleon collision form a precursor
state $|\Phi_{a b} \rangle$ with an average energy and a width of
energy.  As determined by the Feynman production amplitude, the
precursor state is a coherent linear combination of states of various
$CJLS$ states. The probability amplitude for the production of the
bound $CnJLS$ state is then obtained as the projection of this
precursor state to this bound state or to the combination of this
bound state with a soft gluon, after taking into account the evolution
of the precursor state.

We can express this mathematically as follows. The initial precursor
state $\Phi_{a b}$ of the $Q$-$\bar Q$ pair from the collision at
$t_i=0$ is represented by the state vector
\begin{eqnarray}
|\Phi_{a b}(t_i) >=
{\cal M}(ab \rightarrow Q(P/2+q){\bar Q}(P/2-q))
 |Q(P/2+q){\bar Q}(P/2-q)>
\end{eqnarray}
where ${\cal M}(ab \rightarrow Q(P/2+q){\bar Q}(P/2-q))$ is the
Feynman amplitude for the $a+b \rightarrow Q + \bar Q$ process, and
$P$ and $q$ are the center-of-mass and relative momentum of $Q$ and
$\bar Q$.
One can perform a decomposition in terms of color and angular momentum
states as
\begin{eqnarray}
\label{eq:adm}
|\Phi_{a b}(t_i) > 
=\sum_{C J L S }
{\tilde \phi}^C_{JLS}( q)   |Q {\bar Q}[{}^SL_J^C](P)> .
\end{eqnarray}
A bound $Q\bar Q$ state with quantum numbers $JLS$ can be written as
\begin{eqnarray}
\label{eq:c1}
|\Psi_{nJLS};Pq>= \sqrt{2M_{nJLS} \over 4m_Q m_{\bar Q}}
{\tilde R}_{nJLS}({ q}) |Q {\bar Q}[{}^SL_J^{(1)}](P)>.
\end{eqnarray}
In lowest-order perturbative QCD, the probability amplitude for the
direct production of $\Psi_{nJLS}$ is obtained by projecting
$\Phi_{ab} (t_i) $ onto $\Psi_{nJLS}$.  The projection is simplest in
the $Q$-$\bar Q$ center-of-mass system where $P=(M_{nJLS},{\bf 0})$
and $q=(0,{\bf q})$, and the probability amplitude is
\cite{Pes95,Cra91}
\vskip -0.4cm
\begin{eqnarray}
\label{eq:overlap}
<\Psi_{nJLS};Pq|\Phi_{a b}(t_i) >
= \sqrt{ 2M_{nJLS} \over 4m_Q m_{\bar
            Q}} \int {d^3{\bf q} \over (2\pi)^3} {\tilde R}_{nJLS}({\bf
            q}) {\tilde \phi}_{JLS}^{(1)}({\bf q}).
\end{eqnarray}
\vskip -.2cm
\noindent
Because the bound state $\Psi_{nJLS}$ is a color-singlet state, the
above projection will involve only color-singlet components of the
admixture in Eq. (\ref{eq:adm}).

In the next-order perturbation theory, the color-singlet bound state
$\Psi_{nJLS}$, accompanied by a soft gluon $g_s$, can be produced by
the color-octet component of $ \Phi_{ab}$ in Eq.\ ({\ref{eq:adm}).
The probability amplitude for the production of the bound state
$\Psi_{nJLS}$ accompanied by a soft gluon $g_s$ at the hadronization
time $t$ is
\begin{eqnarray}
<[\Psi_{nJLS};Pq]g_s|
\Phi_{ab}(t)>=<[\Psi_{nJLS};Pq]g_s|U(t,t_i)| \Phi_{ab}(t_i)>,
\end{eqnarray}
where $U(t,t_i)$ is the evolution operator.

The state $\Psi_{nJLS}$ can also be produced indirectly through the
production of different bound states $\Psi_{n'J'L'S'}$ which
subsequently decay into $\Psi_{nJLS}$.  The production probability,
including direct, indirect, and color-octet contributions, is then the
sum of the absolute squares of various amplitudes.  Heavy quarkonia
can be produced by different parton combinations such as $g$-$g$,
$q$-$\bar q$, and $g$-$q$ collisions, which will lead to different
precursor states.  The total production probability will be the sum
from all precursor states.

\vspace*{-0.4cm}
\section{Propagation of a Coherent Precursor in Nuclear Matter} 
\vspace*{-0.1cm}

A coherent precursor propagates through the nuclear medium as a single
object with a single absorption cross section. The time of evolution
$t$ can be represented equivalently by the corresponding path length
$L/v$, where $v$ is the velocity of the precursor in the medium.  The
state vector after propagating a distance $L$ in the nuclear medium is
related to the state vector after production by
\begin{eqnarray}
\label{eq:rat}
|\Phi_{ab}(L) \rangle  = e^{-\rho \sigma_{abs}(ab) L/2}|\Phi_{ab}(L=0)
\rangle\,,
\end{eqnarray}
where $\sigma_{abs}(ab)$ is the precursor absorption cross section for
the collision of the precursor $\Phi_{ab}$ with a nucleon, and $\rho$
is the nuclear matter number density.

When we include precursors from different parton collisions leading to
the production of the bound state $\Psi_{nJLS}$, there will be
different factors $e^{-\rho \sigma_{abs}(ab) L/2}$ for different
parton combinations $a$-$b$.  At fixed-target energies, where the
total yield of $J/\psi$ in the forward direction is dominanted by
contributions from $gg$ collisions, there is essentially only a single
survival factor for the total yield in forward directions.  The ratio
of the production of various bound states in a $pA$ collision to a
$pp$ collision will be independent of the mass number of the nucleus,
as the production of all different bound states comes from the
projection of the precursor state onto the bound states after the
absorption.  As a consequence of Eq.\ (\ref{eq:rat}) we have
\begin{eqnarray}
{\sigma(pA\rightarrow \psi'~X) \over \sigma(pp\rightarrow \psi'~X)  }
={ \sigma(pA\rightarrow \psi~X)  \over \sigma(pp\rightarrow \psi~X) }
={ \sigma(pA\rightarrow \chi~X)  \over \sigma(pp\rightarrow \chi~X) }
.
\end{eqnarray}
Experimentally, the ratio $\psi'/(J/\psi)$ is observed to be
approximately a constant of the atomic numbers \cite{Ald91}, in
agreement with the present picture.  The present picture predicts
further that the ratio of the $\chi$ yield to the $J/\psi$ yield will
also be independent of the mass number in $pA$ collisions.  It will be
of interest to test such a prediction in the future.

Recently, C.\ W.\ Wong \cite{Woncw98} pointed out that because of the
nature of the two-gluon coupling model of the Pomeron, channel
coupling between the color-octet and singlet states are weak, and the
propagation of an admixture of color states may nonetheless show up as
two different absorption components even for a coherent admixture of
color states.  It is important to have accurate measurements of the
absorption curve to separate out the different rates of absorption of
the different components.

We conclude this section by remarking that the available data points
are sparse and have large uncertainties.  It is difficult to separate
out the different components of the absorption process.  It will be of
great interest to perform accurate measurements of $J/\psi$ production
in $pA$ collisions for a large set of nuclei so as to infer from the
survival probability the number of incoherent components and their
different absorption cross sections in order to obtain a better
description of the $J/\psi$ precursor.

\vskip -0.3cm
\section*{Acknowledgments}
\vskip -0.1cm

The author would like to thank Drs. T. Barnes and C. W. Wong for
helpful discussions.  This research was supported by the Division of
Nuclear Physics, U.S.D.O.E.  under Contract No. DE-AC05-96OR21400
managed by Lockheed Martin Energy Research Corp..


\begin{thebibliography}{99}

\bibitem{Mat86} T. Matsui and H. Satz, Phys. Lett. B178, 416 (1986).

\bibitem{Gon96} M. Gonin, NA50 Collaboration, Nucl. Phys. 
A610, 404c (1996).

\bibitem{Lou96} C. Louren\c co, NA50 Collaboration, Nucl. Phys.  A610,
552c (1996).

\bibitem{Ein75} M. B. Einhorn and S. D. Ellis, Phys. Rev. D12, 2007
(1975).

\bibitem{Fri77}
H. Fritzsch, Phys. Lett. B67, 217 (1977).

\bibitem{Cha78} Chang Chao-Hsi, Nucl. Phys. B172, 425 (1980);
E. L. Berger and D. Jones, Phys. Rev. D23, 1521 (1981); R. Baier and
R. R\"uckl, Phys. Lett. B102, 364 (1981).

\bibitem{Bod95}
G. T. Bodwin, E. Braaten, and G. P. Lepage, Phys. Rev. D51, 1125
(1995).

\bibitem{Bra96}
E. Braaten, S. Fleming, and T. C. Yuan,
Ann. Rev. Nucl. Part. Sci. 46,  197 (1996).

\bibitem{Ben96} E. Beneke and I. Z. Rothstein, Phys. Rev. D54, 2005
(1996); erratum $ibid.$ 7082;
M. Beneke, hep-ph/9712298.

\bibitem{Cac97b}
M. Cacciari, hep-ph/9706374.

\bibitem{Pet98}
A. Petrelli, M. Cacciari, M. Greco, F. Maltoni, and M. L. Magano,
Nucl. Phys. B514, 245 (1998).

\bibitem{Kha96} D. Kharzeev, Nucl. Phys. A610, 418c (1996);
D. Kharzeev, hep-ph/9802037.

\bibitem{Won96} C. Y. Wong, Phys. Rev. Lett. 76, 196 (1996);
C. Y. Wong, Phys. Rev. C55, 2621 (1997);
C. Y. Wong, Nucl. Phys. A630, 487 (1998).

\bibitem{Won97b}
 C. Y. Wong, Chin. J. Phys. 35, 857 (1997)
(HEP-PH 9712320).

\bibitem{Won98a}
 C. Y.  Wong and C. W. Wong,
Phys. Rev. D57, 1838 (1998).

\bibitem{Van95}
M. V\"antinen, P. Hoyer, S. J. Brodsky, and W. K. Tang,
Phys. Rev. D51, 3332 (1995).

\bibitem{Abe97}
F. Abe $et~al.$, Phys. Rev. lett. 79, 572 (1997).

\bibitem{Ant93}
L. Antoniazzi $et~al.$, Phys. rev. Lett. 70, 383 (1993).

\bibitem{Bin87}
C. Biino $et~al.$, Phys. Rev. Lett. 58, 2523 (1987).

\bibitem{Ake93}
C. Akerlof $et~al.$, E537 Collaboration, Phys. Rev. D48, 5067 (1993).

\bibitem{Cho96}
P. Cho and A. K. Leibovich, Phys. Rev. D53, 150 (1996); 
P. Cho and A. K. Leibovich, Phys. Rev. D53, 6203 (1996).

\bibitem{Cac97a} 
M. Cacciari and  M. Kr\" amer,
Phys. Rev. Lett. 76,  4128 (1996).

\bibitem{Bla52} J. M. Blatt and V. F. Weisskopf, {\it Theoretical
Nuclear Physics}, John Wiley and Sons, New York, 1952.

\bibitem{Tan96}
W. K. Tang 
and M. V\"antinen,
Phys. Rev. D54, 4349 (1996).

\bibitem{Won98b}
C.\ Y.\ Wong, ORNL Preprint ORNL-CTP-9804 (1998).

\bibitem{Bar93} T. Barnes,  Proceedings of Third-Workshop on Tau-Charm
Factory, Marbella, Spain, June 1993, p. 411.

\bibitem{Low75}
F. E. Low, Phys. Rev. D12, 163 (1975);
S. Nussinov, Phys. Rev. Lett. 34, 1286 (1975).

\bibitem{Dol92}
J. Dolej\v s\' i and J. H\"ufner, Z. Phys. C54, 489 (1992).

\bibitem{Woncw96b}
C. W. Wong, Phys. Rev. D{54},   R4199  (1996).

\bibitem{And77}
R. L. Anderson $et~al.$, Phys. Rev. Lett. 38, 263 (1977);
C. Gerschel and J. H\"ufner, Phys. Lett. B207, 253 (1988).

\bibitem{Pes95}
M. E. Peskin and D. V. Schroeder, {\sl An Introduction to Quantum
Field Theory}, Addision Wesley Publishing Company, 1995.

\bibitem{Cra91}
H. W. Crater, Phys. Rev. A44, 7065 (1991).

\bibitem{Qia97}
C. F. Qiao, X. F. Zhang, and W. Q. Chao, hep-ph/9708258.

\bibitem{Ald91}
D. M. Alde $et~al.$, Phys. Rev. Lett. 66, 133 (1991).

\bibitem{Woncw98}
C. W. Wong, Phys. Rev. D58, 037501 (1998). 

\end{thebibliography}
\end{document}